\documentclass[reprint,amsmath,amssymb,aps,pre]{revtex4-2}
\usepackage{graphicx}
\usepackage{dcolumn}
\usepackage{bm}
\usepackage[utf8]{inputenc}
\usepackage[T1]{fontenc}
\usepackage{physics}
\usepackage{amsmath}
\usepackage{hyperref}
\usepackage{placeins}
\usepackage{float}
\usepackage{color}
\usepackage{xr}
\usepackage[most]{tcolorbox}
\usepackage{tikz}
\usepackage{dcolumn}
\usepackage{relsize}
\usepackage{parskip}
\usepackage[normalem]{ulem}

\begin{document}
\title{Vortex triplets, symmetry breaking, and emergent nonequilibrium plastic crystals in an active-spinner fluid}
\author{Biswajit Maji\textsuperscript{1}}
\email{biswajitmaji@iisc.ac.in}
\author{Nadia Bihari Padhan\textsuperscript{2}}
\email{nadia_bihari.padhan@tu-dresden.de}
\author{Rahul Pandit\textsuperscript{1}}
\email{rahul@iisc.ac.in}
\affiliation{
\textsuperscript{1}Centre for Condensed Matter Theory, Department of Physics, Indian Institute of Science, Bangalore 560012, India \\
\textsuperscript{2}Institute of Scientific Computing, TU Dresden, 01069 Dresden, Germany
}




\begin{abstract}
{The formation of patterns and exotic nonequilibrium steady states in active-fluid systems continue to pose challenging problems -- theoretical, numerical, and experimental -- for statistical physicists and fluid dynamicists. We combine theoretical ideas from statistical mechanics and fluid mechanics to uncover a new type of self-assembled crystal of vortex triplets in an active-spinner fluid. 
We begin with the two-dimensional Cahn–Hilliard–Navier–Stokes (CHNS) model for a binary-fluid system of active rotors that has two important ingredients: a scalar order parameter field $\phi$ that distinguishes regions with clockwise ($CW$) and counter-clockwise ($CCW$) spinners; and an incompressible velocity field $\bm u$. In addition to the conventional CHNS coupling between $\phi$ and $\bm u$, this model has a \textit{torque-induced activity term}, with coeffcient $\tau$, whose consequences we explore. We demonstrate that, if we increase the activity $\tau$, it overcomes dissipation and 
this system displays a hitherto unanticipated emergent triangular crystal, with spinning vortex triplets at its vertices. We show that this is a nonequilibrium counterpart of an equilibrium plastic crystal.
We characterise the statistical properties of this novel crystal
and suggest possible experimental realisations of this new state of active matter.}

\end{abstract}
\maketitle
\section*{Emergent vortex triplets in an active-spinner fluid}
\begin{figure*}[htp]
\includegraphics[width=1.0\textwidth]{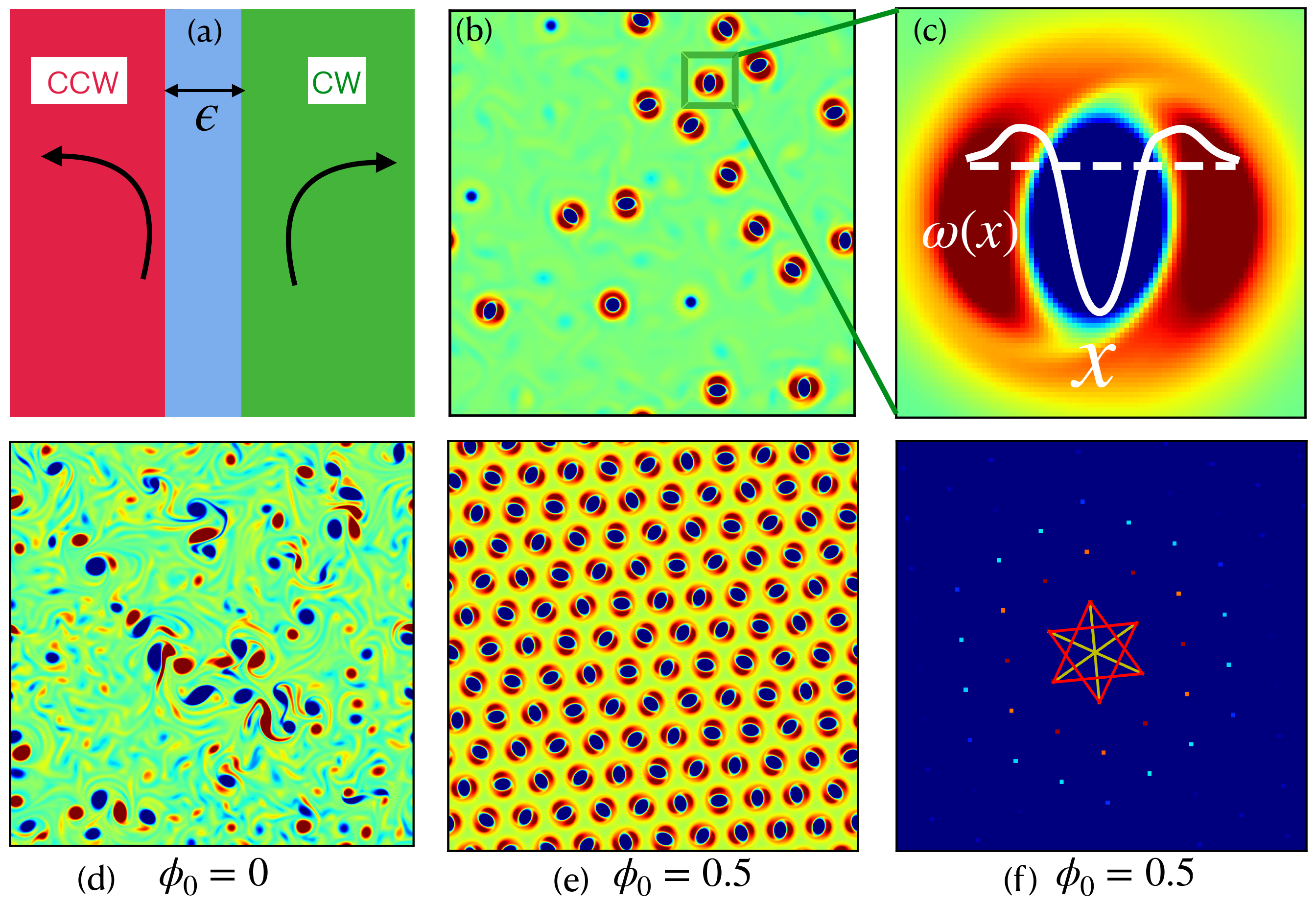}  
\caption{\textbf{Illustrations of vortex triplets and their crystals:} (a) A schematic diagram showing $CCW$ (red) and $CW$ (green) phases separated by an interfacial region (blue) of width $\epsilon$. (b) A pseudocolor plot of the vorticity, $ \omega(x, y, t)$, illustrating a vortex triplet at  $\phi_0 = 0.5$ during the transient state. (c) A zoomed-in view of the vorticity field, $ \omega(x,y,t)$, highlights the emergence of a vortex triplet at $\phi_0 = 0.5 $ and $ \tau = 4$; the white curve represents a one-dimensional profile extracted from the two-dimensional vorticity field, demonstrating the structure of a single vortex triplet. (d) Pseudocolor plot of the vorticity $\omega(x, y, t)$, showing vortex doublets [cf. Ref.~\cite{maji2025emergent}] at $\phi_0 = 0.0$. (e) A pseudocolor plot of the vorticity field $ \omega(x, y, t) $ unveils the formation of a stable triangular crystal at $\phi_0 = 0.5 $, see video V0 in the Supplemental Information. (f) Pseudocolor plots illustrate the Fourier transformation of the vorticity field for a triangular triplet vortex crystal ($\omega(x, y, t)$) at $\phi_0=0.5$, $\sigma=1$, and $\tau = 4.0$.} 
\label{fig:schematic}
\end{figure*} 
Interest in active materials and active fluids continues to grow apace~\cite{maji2025emergent,padhan2025cahn,te2024review,bowick2022symmetry,shankar2022topological,marchetti2013hydrodynamics,pandit2025particles}. Such systems often exhibit nonequilibrium statistically steady states (NESSs) because they contain active entities, generically referred to as particles, that transform internally supplied energy into mechanical work~\cite{marchetti2013hydrodynamics,ramaswamy2010mechanics}. These NESSs display, \textit{inter alia}, emergent patterns, as witnessed, e.g., in crowds ~\cite{castellano2009statistical,bottinelli2016emergent}, fish schools ~\cite{becco2006experimental}, bird flocks ~\cite{bialek2012statistical,cavagna2010scale}, and bacterial colonies~\cite{chen2017weak}.
Most research on active suspensions has concentrated on particles that undergo \textit{translational motion}, such as bacteria and their non-living counterparts~\cite{aranson2013active,elgeti2015physics,driscoll2019leveraging}; but, of late, there has been growing interest in systems with \textit{self-rotating} particles~\cite{maji2025emergent,yeo2015collective,goto2015purely,nguyen2014emergent,kokot2017active,banerjee2017odd}. Rotational degrees of freedom enrich the palette of choices for self-organised NESSs in active fluids as we show below.

Even if an isolated micro-rotor lacks self-motility, an assembly of such rotors, when interconnected, has the potential for motility. The initial observation of self-organization in externally driven rotors was made by Grzybowski, Stone, and Whitesides~\cite{grzybowski2000dynamic}, who found dynamic patterns of millimetre-sized magnetic disks at a liquid-air interface, subjected to a magnetic field produced by a rotating permanent magnet. Subsequently, rotors have generated intriguing NESSs in biological systems, such as rotational motion in uniflagellar mutants of \textit{Chlamydomonas}~\cite{huang1982uniflagellar,brokaw1982analysis}, active two-dimensional crystals of rotating cells~\cite{petroff2015fast} in suspensions of the bacterium \textit{Thiovulum majus}, spinning organisms like sperm-cell clusters~\cite{riedel2005self}, dancing \textit{Volvox} algae~\cite{drescher2009dancing}, and starfish embryos~\cite{tan2022odd}. In addition, there is a growing array of synthetic rotors whose motion is driven by chemical reactions~\cite{wang2009dynamic} or external stimuli such as magnetic fields ~\cite{zhang2020reconfigurable,grzybowski2001dynamic,kokot2017active,han2020reconfigurable}, electric fields ~\cite{shields2018supercolloidal}, oscillating platforms~\cite{tsai2005chiral}, ultrasound~\cite{sabrina2018shape}, optical ﬁelds~\cite{friese1998optical}, and applied airflow~\cite{farhadi2018dynamics}. Agent-based simulations, which do not include hydrodynamics, have revealed, for gear-shaped spinners~\cite{spellings2015shape} and rotating dimers ~\cite{van2016spatiotemporal}, that a system of clockwise ($CW$) and counter-clockwise ($CCW$) rotating spinners segregates into $CW$-rich and $CCW$-rich phases. Indeed, the self-assembly of rotors is a common characteristic of aggregating spinners~\cite{schwarz2012phase} and the self-organization of rotating molecular motors in membranes~\cite{lenz2003membranes}.

Nonlinear hydrodynamic interactions play vital roles in the self-organization of systems of rotating disks~\cite{grzybowski2000dynamic,grzybowski2001dynamic}, so it is important to go beyond the Stokes approximation [cf. the self-organization of vortices, at high Reynolds numbers, into vortex crystals~\cite{aref2002vortex,durkin2000experiments,perlekar2010turbulence,gotze2011dynamic,gupta2017melting}]. With this in mind, we use the minimal hydrodynamical description~\cite{sabrina2015coarsening} for $CCW$ and $CW$ spinners, namely, the active-rotor (or active-spinner) Cahn-Hilliard-Navier-Stokes (ARCHNS) model~\cite{sabrina2015coarsening,maji2025emergent} that is defined by the following partial differential equations (PDEs):
\begin{eqnarray}
    \partial_t \phi + (\bm u \cdot \nabla) \phi &=& M \nabla^2 \left( \frac{\delta \mathcal F}{\delta \phi}\right)\,; \label{eq:phi}  \\
    \mathcal F[\phi,\nabla \phi] &=& \int_{\Omega}\left[\frac{3}{16} \frac{\sigma}{\epsilon}(\phi^2-1)^2 + \frac{3}{4}\sigma\epsilon|\nabla \phi|^2\right];\label{eq:functional} \\
     \partial_t \bm u + (\bm u \cdot \nabla) \bm u &=& -\nabla p + \nu \nabla^2 \bm u -\frac{3}{2}\sigma\epsilon (\nabla^2\phi\nabla\phi) \nonumber \\    
     &+& \nabla\times\mathcal(\bm \tau\phi)-\beta\bm u;\label{eq:vel} \\
    \nabla \cdot \bm u &=& 0\,; \quad \bm \omega = (\nabla \times \bm u)\,;\label{eq:incom} 
    \end{eqnarray}
here, $M$ is the mobility coefficient,  $\epsilon$ the interfacial width, $\sigma$ the surface tension, $\nu$ the kinematic viscosity, $\beta$ the friction, and $\tau$ the magnitude of the active torque $\bm{\tau}$. The fluid is incompressible [Eq.~\eqref{eq:incom}], so we work with a constant unit density. The scalar field $\phi(x,y,t)$, which is positive (negative) in $CCW$ ($CW$) regions, plays a crucial role in the \textit{activity} of the system via the term $\nabla\times\mathcal(\bm \tau\phi)$; furthermore, this field is also active in the fluid-dynamics sense~\cite{padhan2025cahn} because $\phi$ acts back on the advecting fluid velocity $\bm u$  [see Model and Methods for details]. 
Our initial conditions are $\omega(x,y,t=0) = 0$ and $\phi(x,y,t=0) = \phi_0 + \delta\phi(x,y,t=0)$, with $\phi_0$ a space and time-independent constant and $\delta\phi(x,y,t=0)$
independent random numbers distributed uniformly in $[-0.1, 0.1]$; as we show below, $\phi_0$ is an important control parameter.

Our study of this ARCHNS model helps us to uncover intriguing vortex triplets, symmetry breaking, and emergent nonequilibrium plastic crystals in a fluid of counter-rotating active spinners.
The schematic diagram in Fig.~\ref{fig:schematic} (a) shows $CCW$  (red) and $CW$ (green) regions separated by an interface (blue) in an active-spinner fluid. The ARCHNS model for such a fluid can give rise to vortex triplets, which are apparent in the pseudocolor plot of the vorticity $\omega$ in Fig.~\ref{fig:schematic} (b) [one triplet is shown enlarged in Fig.~\ref{fig:schematic} (c)]. Several control parameters affect the emergence of vortex triplets and their crystalline assemblies. These are $\phi_0$, the mean concentration difference between $CW$ and $CCW$ spinners, the magnitude  $\tau$ of the active torque, the surface tension $\sigma$, the kinematic viscosity $\nu$, 
the interface width $\epsilon$, and the friction $\beta$, which can be combined to obtain the dimensionless Reynolds number $Re$, the Cahn number $Cn$, the P\'eclet number $Pe$, and the Weber number $We$ that we discuss in the Section on Models and Methods [see Table~\ref{tab:param}].
\begin{figure*}[htp]
\includegraphics[width=1.0\textwidth]{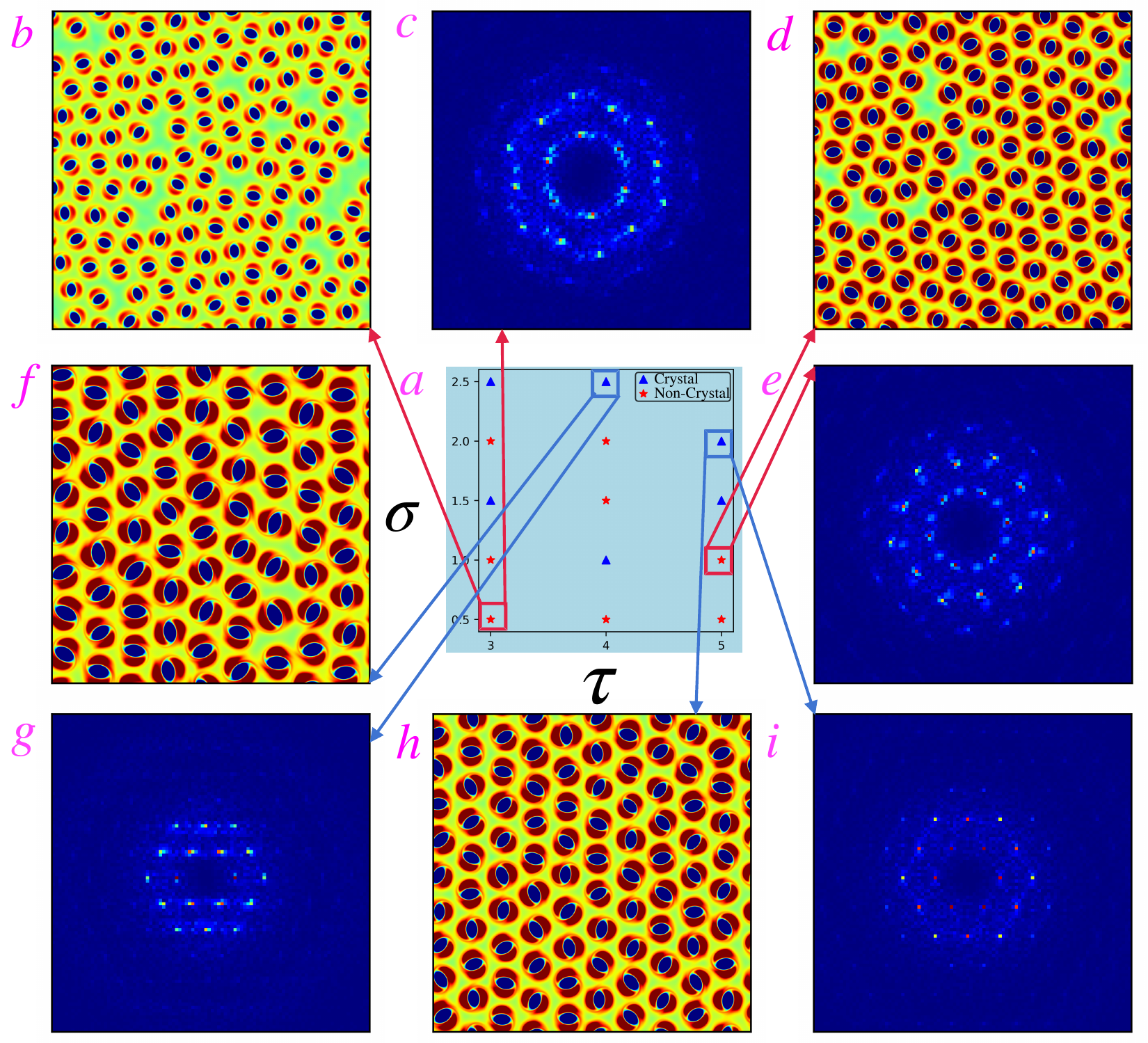}  
\caption{{\bf{Nonequilibrium phases, transitions, and morphologies for $\phi_0 = 0.5$ and representative values in the $\tau-\sigma$ plane:}} (a) Nonequilibrium phase diagram in the $\tau-\sigma$ plane; blue triangles indicate triplet-vortex crystals and red stars disordered crystals. Illustrative pseudocolor plots of the vorticity field $\omega(x, y, t)$ along with their associated spatial Fourier transforms, which yield signatures of the crystalline ordering in reciprocal space for: (b) and (c) 
$\tau = 3$ and $\sigma = 0.5$; (d) and (e) $\tau = 5$ and $\sigma = 1$; (f) and (g) $\tau = 4$ and $\sigma = 2.5$; (h) and (i) $\tau = 5$ and $\sigma = 1.5$. For the complete spatiotemporal evolution of these patterns, see videos V1-V4 in the Supplemental Information.}
\label{fig:vort_diagram}
\end{figure*}
\begin{figure*}[htp]
\includegraphics[width=1.0\textwidth]{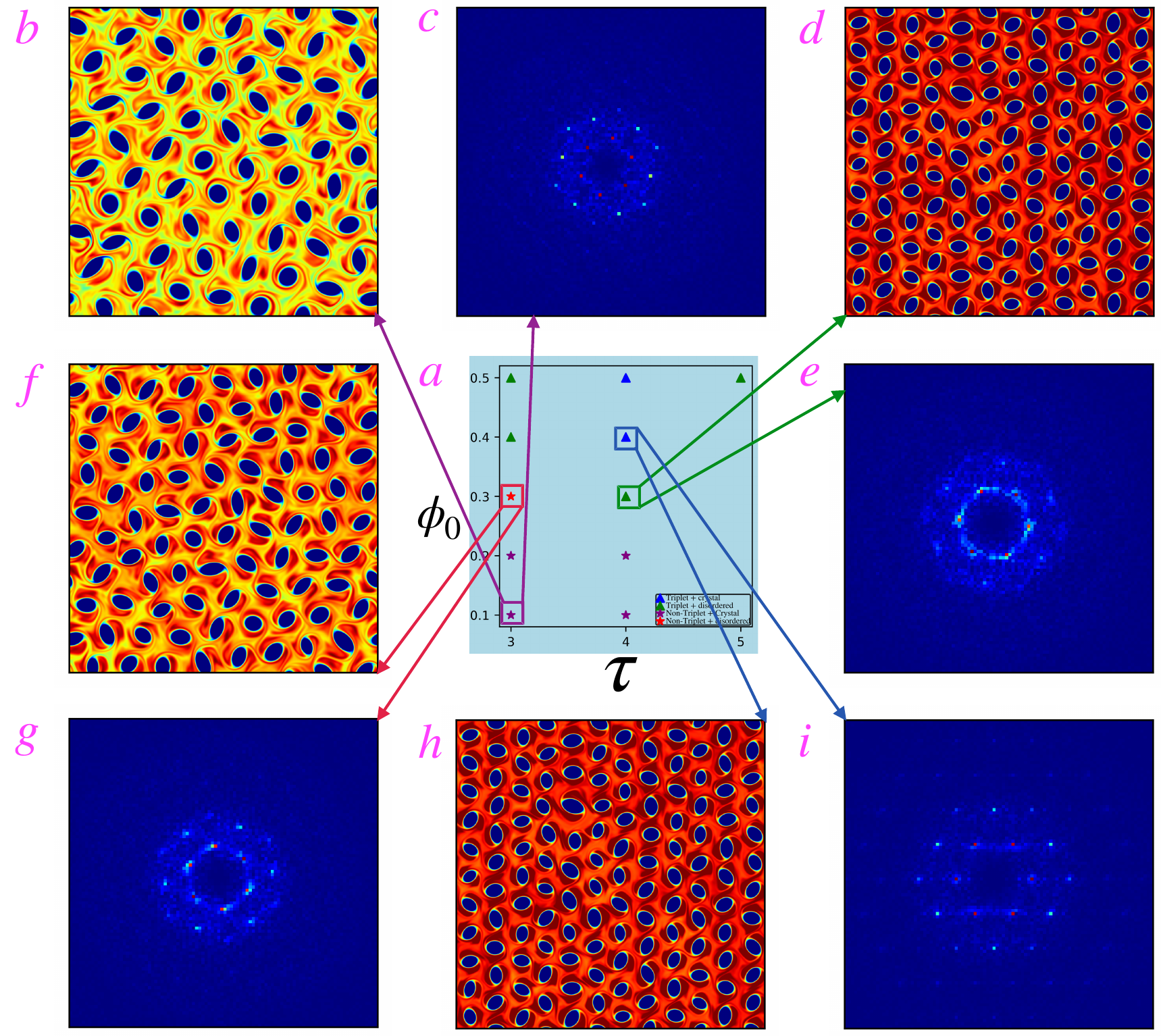}  
\caption{{ \bf{Nonequilibrium phases, transitions, and morphologies for $\sigma = 1$ and representative values in the $\phi_0-\tau$ plane:}} (a) Nonequilibrium phase diagram in the $\phi_0-\tau$ plane; blue triangles indicate triplet-vortex crystals, green triangles triplet disordered crystals, purple stars non-triplet crystals, and red stars non-triplet disordered crystals. Illustrative pseudocolor plots of the vorticity field $\omega(x, y, t)$ along with their associated spatial Fourier transforms, which yield signatures of the crystalline ordering in reciprocal space for: (b) and (c) 
$\tau = 3$ and $\phi_0 = 0.1$; (d) and (e) $\tau = 4$ and $\phi_0 = 0.3$; (f) and (g) $\tau = 3$ and $\phi_0 = 0.3$; (h) and (i) $\tau = 4$ and $\phi_0 = 0.4$. For the complete spatiotemporal evolution of these patterns, see videos V5-V8 in the Supplemental Information.}
\label{fig:field_diagram}
\end{figure*}
\begin{figure*}[htp]
     \includegraphics[width=1.0\textwidth ]{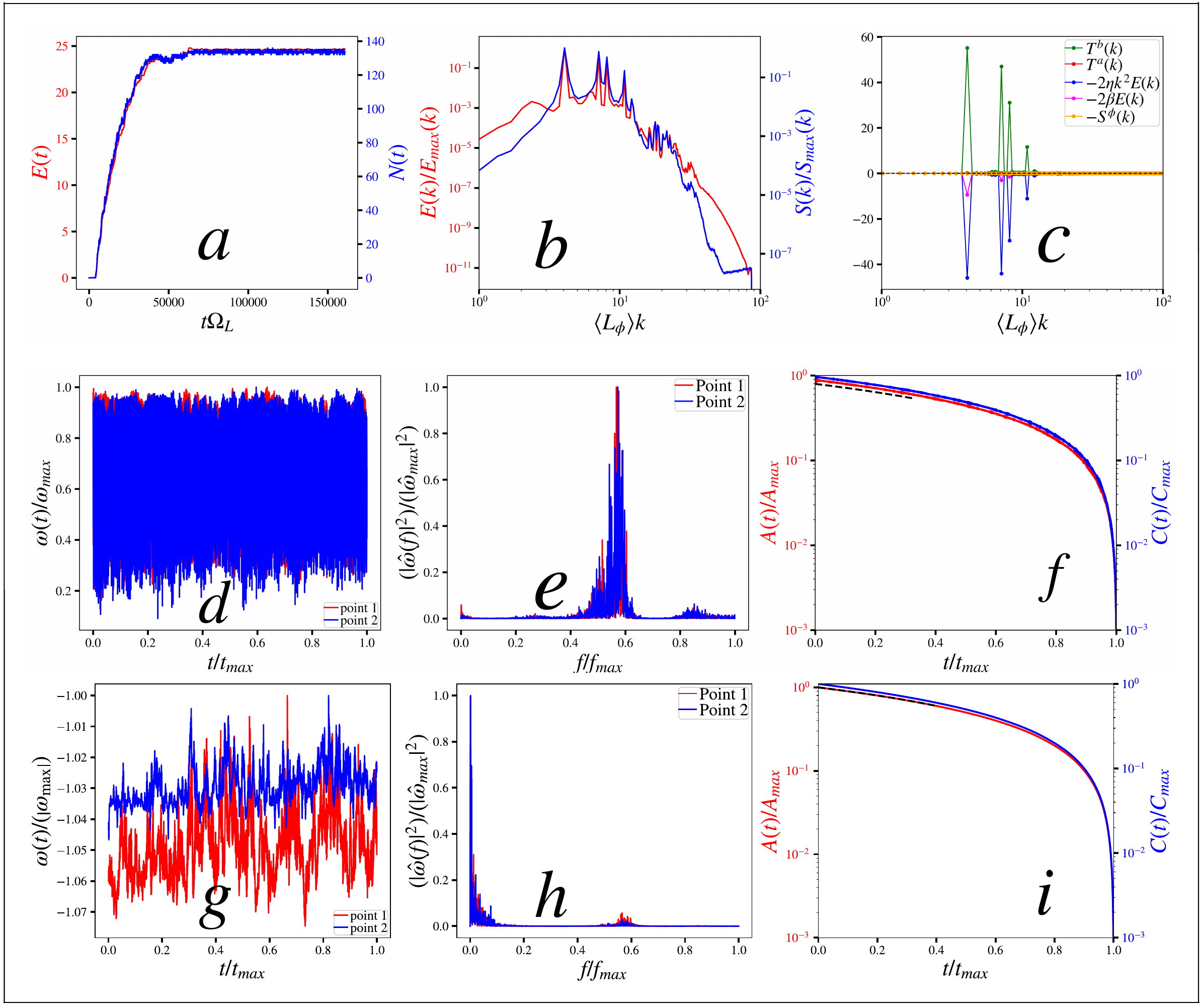} 
    \caption{{\bf{Plots for the illustrative parameter values  $\tau=4$, $\sigma=1$, and $\phi _{0}=0.5$:}} (a) The fluid energy $E(t)$ (red) and the number $N(t)$ (blue) of vortex triplets versus the scaled time $t\Omega_L$, where $\Omega_L$ is the integral-scale frequency. 
    (b) Log-log plots against the scaled wavenumber $\langle L_\phi \rangle k$ of the scaled fluid-energy spectrum $E(k)/E_{max}(k)$ (red)
    and the scaled phase-field spectrum $S(k)/S_{max}(k)$ (blue); the sharp Bragg peaks at $ k = k_0  \simeq 2\pi/a$, with the lattice spacing $a \simeq 0.5842$, $k= \sqrt{3}k_0$, and $k = 2k_0$ are associated with the triplet-vortex crystal [Figs.~\ref{fig:schematic} (e) and (f)]. 
    (c) Semilog plots versus $\langle L_\phi \rangle k$ of $T^a(k)$ (red), $T^{b}(k)$ (green), $S^{\phi}(k)$ (orange), $2\nu k^2 E(k)$ (dark-blue), and $2\beta E(k)$ (magenta) [see Eqs.~\eqref{eq:Sk} and \eqref{eq:Tketc}]. 
    (d) The scaled vorticity  $\omega(t)/\omega_{max}$ versus the scaled time $t/t_{max}$, where $t_{max}$ is the time for which we collect data in the nonequilibrium plastic crystalline (PC) state; we present plots, in red and blue, for two neighboring vortex triplets, with $\omega(t)$ from points to the right of the triplet centre (at a distance of $\simeq 0.2731 a$). 
    (e) Plots versus the scaled frequency $f/f_{max}$ of the power spectra (red and blue) of the two time series in (d) [their means have been subtracted and carets denote the temporal Fourier transform]; both these power spectra indicate 
    temporal quasiperiodicity, because the principal peaks can be labelled as $(n_1f_1+n_2f_2)$, where the integers $n_1,\,n_2 \in [\ldots -2, -1, 0, 1, 2, \ldots]$ and the fundamental frequencies $f_1$ and $f_2$ are incommensurate, i.e., $f_1/f_2$ is an irrational number; these peaks rise out of a broad-band background, a signature of temporal chaos that leads to the exponential decay [see (f)] of temporal correlation functions.  (f) Plots versus $t/t_{max}$  of the vorticity autocorrelation function $A(t)$ and the vorticity cross-correlation $C(t)$ [Eq.~\eqref{eq:timecorr}], illustrating an exponential decay (indicated by the dashed black line). The plots in (g), (h), and (i) are the counterparts of those in (e), (f), and (g), but with $\omega(t)$ from a point near the centre of a triplet.}
\label{fig:energy}    
\end{figure*}
\section*{The self-assembly of vortex triplets into a crystal: spontaneous symmetry breaking}

By tuning these parameters, we find remarkable and unanticipated nonequilibrium states.
These include the turbulent state~\cite{maji2025emergent} depicted in Fig.~\ref{fig:schematic} (d), for $\phi_0=0$ and $\tau = 4$, and, on increasing $\phi_0$ to $0.5$, the emergent triangular crystal of vortex triplets shown by the real-space 
pseudocolor plot of $\omega$ in Fig.~\ref{fig:schematic} (e) and its spatial Fourier transform in Fig.~\ref{fig:schematic} (f), which exhibits clearly the reciprocal lattice~\cite{Ashcroft76,solyom2007fundamentals} for a triangular crystal. Video V0 [in the Supplemental Information], illustrates the spatiotemporal evolution of this crystal and shows explicitly that the vortex triplets spin rapidly, but out-of-phase, so that there is no net vorticity~\footnote{In our pseudospectral DNS, we work with zero mean velocity and zero mean vorticity.}. Thus, we have uncovered, for the first time, a \textit{nonequilibrium-plastic-crystalline} state [Fig.~\ref{fig:schematic} (d) and Video V0]. We recall that \textit{equilibrium} plastic crystals are found in mesogenic systems; such plastic crystals display \textit{positional crystalline order}, but \textit{no orientational order}~\cite{mahato1988liquid,sherwood1979plastically} as in the plastic-crystalline phase of methane~\cite{bini1997high}.

\textit{Spontaneous symmetry breaking} leads to the formation of this nonequilibrium triangular plastic crystal. Therefore, this crystal is \textit{qualitatively different} from vortex crystals or cellular flows, which are formed by the imposition of a force that is periodic in space [see, e.g., Refs.~\cite{gotoh1984instability,braun1997bifurcations,ouellette2007curvature,perlekar2010turbulence,gupta2017melting}] that is an \textit{external symmetry-breaking field} in the terminology of condensed-matter field theory [see, e.g., Refs.~\cite{altland2010condensed,fradkin2013field}].

\begin{figure*}[htp]
   \includegraphics[width=1.0\linewidth]{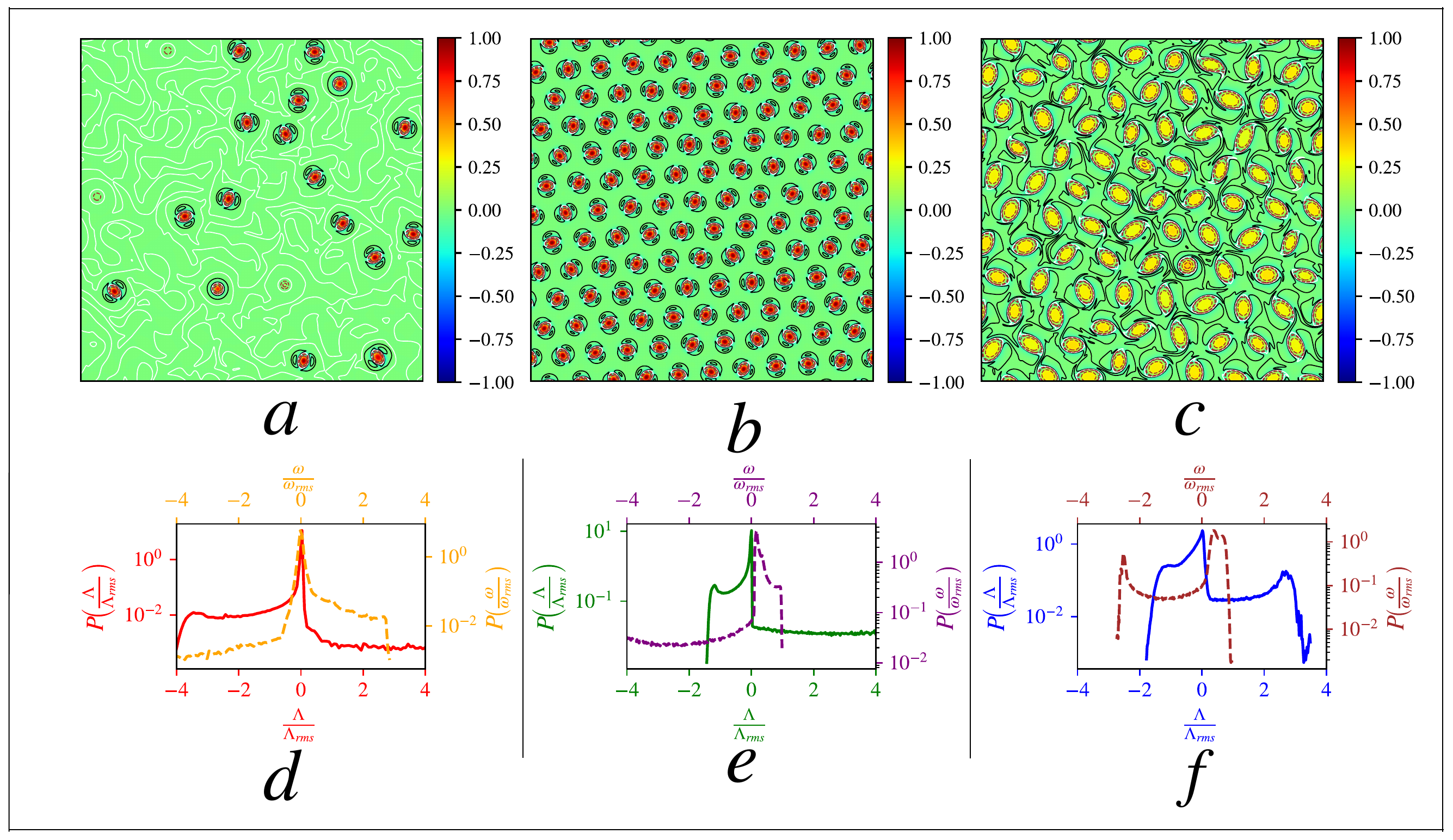}
    \caption{\textbf{Pseudocolour plots of the normalised Okubo–Weiss parameter $\Lambda / \Lambda_{\max}$:} (a)  $\tau = 4$ and $\sigma = 0.5$ [the early-time transient in Fig.~\ref{fig:schematic} (b)], (b)  $\tau = 4$ and  $\sigma = 0.5$ [the statistically steady crystal of vortex triplets in  Fig.~\ref{fig:schematic} (e)], and (c) $\tau = 3$ and $\sigma = 0.3$ [the fluctuating crystal of incipient vortex triplets in Fig.~\ref{fig:field_diagram} (e)];
    (a), (b), and (c) also show overlaid contours for three different values of $\omega/\omega_{max}$, namely, $0.6$ (black contours), $0$ (white contours), and $-0.6$ (dashed-brown contours). Plots of the probability distribution functions (PDFs) of $\Lambda/\Lambda_{rms}$ and $\omega/\omega_{rms}$ are shown in (d), (e), and (f), corresponding, respectively, to (a), (b), and (c).
    }
\label{fig:Lambda}
\end{figure*}
\section*{Characterising the triangular vortex-triplet crystal}

We now explore the dependence of these NESSs on the parameters in the ARCHNS model. Consider first the illustrative nonequilibrium phase diagram in the $\tau-\sigma$ plane [Fig.~\ref{fig:vort_diagram} (a)], where blue triangles indicate triplet-vortex crystals and red stars disordered crystals. We present the following representative pseudocolor plots of $\omega(x, y, t)$ along with their associated spatial Fourier transforms in reciprocal space, at a typical time [$\tau = 3$ and $\sigma = 0.5$ in Figs.~\ref{fig:vort_diagram} (b) and (c), $\tau = 5$ and $\sigma = 1$
in  Figs.~\ref{fig:vort_diagram} (d) and (e), $\tau = 4$ and $\sigma = 2.5$ in  Figs.~\ref{fig:vort_diagram} (f) and (g), and $\tau = 5$ and $\sigma = 1.5$ in  Figs.~\ref{fig:vort_diagram} (h) and (i)]. These plots bring out several important features: 
\begin{itemize}
\item The disordering of the vortex-triplet crystal occurs via the formation of vacancies or grain boundaries, to use the language of solid-state physics~\cite{Ashcroft76,solyom2007fundamentals}; indeed, incipient crystalline chunks remain and lead to blurred signatures of a triangular lattice in the reciprocal-space plots in Fig.~\ref{fig:vort_diagram}.
\item Different scans through the parameter space [Fig.~\ref{fig:vort_diagram} (a)] can lead to re-entrance, from, say, a disordered phase to a vortex-triplet-crystal and back; this is reminiscent of the complicated sequence of transitions seen in the melting  of a vortex crystal (or cellular flow) by fluid turbulence~\cite{perlekar2010turbulence}, polymeric turbulence~\cite{gupta2017melting}, or interfacial turbulence~\cite{padhan2024interface}.
\item The pseudocolor plots of of $\omega(x, y, t)$ are not frozen in time, for the vortex triplets at different sites rotate rapidly [see videos V1-V4 in the Supplemental Information].
\end{itemize}

We can also go from one nonequilibrium phase to another by tuning $\phi_0$ as we illustrate in Figs.~\ref{fig:field_diagram} (a)-(i). The nonequilibrium phase diagram of our
ARCHNS system is shown in Fig.~\ref{fig:field_diagram} (a) in the $\phi_0-\tau$ plane, with $\sigma = 1$; here,  blue and green triangles indicate, respectively, triplet-vortex crystals and disordered crystals with vortex triplets. Furthermore, we obtain incipient vortex triplets with triangular crystalline and disordered states, at the parameter values indicated by purple and red stars in Fig.~\ref{fig:field_diagram} (a). The pseudocolor plots, at a representative time, of $\omega(x, y, t)$ and their associated spatial Fourier transforms in reciprocal space, which yield signatures of these types of crystalline and disordered states, are portrayed in Figs.~\ref{fig:field_diagram} (a)-(i). We emphasize that these plots of $\omega(x, y, t)$ are not frozen in time; the vortex triplets or incipient-vortex triplets at different sites rotate rapidly [see the videos V5-V8 in the Supplemental Information]. Complicated spatiotemporal evolution occurs for $\phi_0 =0.3,\, \tau =4$, and $\sigma=1$ [Fig.~\ref{fig:field_diagram} (d) and video V6] and  $\phi_0 =0.4,\, \tau =4$, and $\sigma=1$ [Fig.~\ref{fig:field_diagram} (h) and video V8] as the system tries to move from states dominated by doublets and incipient triplets to those with well-formed triplets.

To quantify the spatiotemporal evolution of our active-spinner system, we present, in Fig.~\ref{fig:energy}, various time series and spectra, for the illustrative parameter values  $\tau=4$, $\sigma=1$, and $\phi _{0}=0.5$. The total fluid energy $E(t)$ (red curve) and the number $N(t)$ (blue curve) of vortex triplets are plotted, in Fig.~\ref{fig:energy} (a), versus the scaled time $t\Omega_L$, where $\Omega_L$ is the integral-scale frequency; clearly, $E(t)$ and $N(t)$ show mild fluctuations about their statistically steady mean values at large times. In Fig.~\ref{fig:energy} (b), we present log-log plots  of the scaled fluid-energy $E(k)/E_{max}(k)$ (red curve) and phase-field $S(k)/S_{max}(k)$ (blue curve) spectra versus the scaled wavenumber $\langle L_\phi \rangle k$; the sharp Bragg peaks at $k = k_0  \simeq 2\pi/a$, with the lattice spacing $a \simeq 0.5842$, $k= \sqrt{3}k_0$, and $k = 2k_0$ [for the triplet-vortex crystal shown in Figs.~\ref{fig:schematic} (e) and (f)]. Figure~\ref{fig:energy} (b) also suggests that $\omega \propto \phi$; at large values of the activity $\tau$, this proportionality follows from a theoretical dominant-balance argument~\cite{maji2025emergent} applied to Eq.~\eqref{eq:spectralbalance}. The terms in the spectral-balance equations~\eqref{eq:Sk}-\eqref{eq:Tketc} are given in the semi-log plots in Fig.~\ref{fig:energy} (c) versus $\langle L_\phi \rangle k$ of $T^a(k)$ (red curve), $T^{b}(k)$ (green curve), $S^{\phi}(k)$ (orange curve), $2\nu k^2 E(k)$ (dark-blue curve), and $2\beta E(k)$ (magenta curve). These plots show clearly that the advective and stress-tensor contributions $T^a(k)$ and $S^{\phi}(k)$ are negligibly small at all wavenumbers $k$, so \textit{there is no significant cascade of energy}, which is reminiscent of the active-nematic turbulence  considered in Ref.~\cite{alert2020universal}. Energy that is injected into the flow by the torque term $T^{b}(k)$ (green curve) is dissipated predominantly at the same value of $k$ by the viscous and friction contributions  $2\nu k^2 E(k)$ (dark-blue curve), and $2\beta E(k)$ (magenta curve).
 

The mean vorticity, averaged over the whole simulation domain, vanishes, because we work in a frame where the mean velocity is zero~\footnote{In Fourier space $\tilde{\bm{\omega}}({\bm{k}}) = i {\bm{k}} \times \tilde{\bm{u}}({\bm{k}})$; we set $\tilde{\bm{u}}({\bm{k}}) = 0$, so there is no mean velocity; therefore, $\tilde{\bm{\omega}}({\bm{k}}) = 0$, and there is no mean vorticity.}. However, individual vortex triplets spin, as shown in the videos V0-V8 in the Supplementary Information. We characterise the spinning of such triplets in Figs.~\ref{fig:energy} (d)-(f). In Fig.~\ref{fig:energy} (d) we plot the local scaled vorticity  $\omega(\bm{x}_*,t)/\omega_{max}$ versus the scaled time $t/t_{max}$, at a representative point $\bm{x}_*$ inside a triplet. [Henceforth, we suppress $\bm{x}_*$ for notational convenience.] Here, $\omega_{max}$ is the maximal value of $\omega(t)$, over the duration $t_{max}$ of its time series; we present plots, in red and blue, for two neighbouring vortex triplets [the point at which we measure $\omega(t)$ lies to the right of the centre of the triplet].  Figure~\ref{fig:energy} (e) displays plots, versus the scaled frequency $f/f_{max}$, of the power spectra of the two time series in Fig.~\ref{fig:energy} (d) [their means have been subtracted and carets denote the temporal Fourier transform], which provide evidence for quasiperiodicity, because the principal peaks can be indexed as $(n_1f_1+n_2f_2)$, with the integers $n_1,\,n_2 \in [\ldots, -2, -1, 0, 1, 2, \ldots]$ and incommensurate fundamental frequencies $f_1$ and $f_2$ [i.e., $f_1/f_2$ is an irrational number]. These peaks rise above a broad-band background, a signature of temporal chaos that leads to the exponential decay [see the dashed-black line Fig.~\ref{fig:energy} (f)] of the vorticity autocorrelation function $A(t)$ and the vorticity cross-correlation function $C(t)$ [Eq.~\eqref{eq:timecorr}], which we plot in Fig.~\ref{fig:energy} (f) versus $t/t_{max}$.
The plots in Figs.~\ref{fig:energy} (g), (h), and (i) are the counterparts of those in Figs.~\ref{fig:energy} (e), (f), and (g), but with $\omega(t)$ from a point near the centre of a triplet, where $\omega(t)$ remains negative for the duration of our simulation.

The Okubo-Weiss parameter~\cite{okubo1970horizontal,weiss1991dynamics,perlekar2009statistically,pandit2017overview}, a common and convenient measure 
of the flow topology, is
\begin{equation}
\Lambda(x,y) \equiv \frac{\bm{\omega}^2(x,y) - \bm{\Sigma}^2(x,y)}{8}\,,
\label{eq:Okubo}
\end{equation}
with $\bm{\Sigma}$ the symmetric part of $\partial_i\bm{u}_j$, the velocity-derivative tensor; in vortical regions of the flow, $\Lambda > 0$; by contrast, $\Lambda < 0$ in the extensional parts of the flow. To characterise the flow topologies in our system, we show pseudocolor plots of $\Lambda$ for three representative flows in Figs.~\ref{fig:Lambda} (a), (b), and (c), which correspond, respectively, to the pseudocolor plots of the vorticity $\omega$ in Fig.~\ref{fig:schematic} (b) [an early-time transient], Fig.~\ref{fig:schematic} (e) [the statistically steady crystal of vortex triplets], and Fig.~\ref{fig:field_diagram} (e) [with incipient vortex triplets]. Furthermore, in Figs.~\ref{fig:Lambda} (a), (b), and (c), we have overlaid three contours for three different vaues of $\omega/\omega_{max}$, namely, $0.6$ (black contours), $0$ (white contours), and $-0.6$ (dashed-brown contours) to illustrate 
how  $\Lambda$ partitions the flow into hyperbolic ($\Lambda<0$, strain-dominated or extensional) and elliptical ($\Lambda>0$, rotation-dominated or vortical) regions. Note, in particular, that when we have a well-formed crystal of vortex triplets [Fig.~\ref{fig:schematic} (e) and Fig.~\ref{fig:Lambda} (b)] there is essentially no flow [so $\Lambda=0$] in the regions outside the vortex triplets; and the inverse-cascade of energy, well-known in conventional 2D fluid turbulence, is suppressed as we have mentioned while discussing Fig.~\ref{fig:energy} (c) [cf. Ref.~\cite{van2022spontaneous,van2024vortex}].  
Below these pseudocolor plots of $\Lambda$, we give probability distribution functions (PDFs) of $\Lambda/\Lambda_{rms}$ and $\omega/\omega_{rms}$ in Figs.~\ref{fig:Lambda} (d), (e), and (f). The PDFs of $\omega/\omega_{rms}$ are reminiscent of those in Ref.~\cite{van2022spontaneous,van2024vortex}, where vortex triplets form  a triangular crystal because of a specific external forcing.

\section*{Vortex-triplet crystals: significance and prospectus}

Active-fluid models have led to a variety of exciting nonequilibrium states, including those that show different types of active turbulence~\cite{alert2020universal,shankar2022topological,mukherjee2023intermittency,padhan2023activity,kiran2023irreversibility,padhan2024novel,padhan2025cahn,kiran2025onset,pandit2025particles}; studies of such NESSs in active-spinner systems are in their infancy~\cite{sabrina2015coarsening,maji2025emergent}. Our work has initiated theoretical studies
of exotic states in active-spinner systems using the ARCHNS model.  We have shown how the ARCHNS model~\cite{sabrina2015coarsening,maji2025emergent} leads to the
 spontaneous formation of vortex triplets, which can then self-assemble to yield a symmetry-broken 
 state that we identify as a nonequilibrium plastic crystal, which had not been anticipated earlier. 
 
 Our extensive DNSs have helped us to obtain representative phase diagrams in the parameter space of the ARCHNS model. These phase diagrams should serve as a guide for developing new experimental systems, of $CW$ and $CCW$ active spinners, that display emergent active crystals of vortex triplets. Crystalline arrays of vortex triplets have been obtained so far only in an externally forced fluid~\cite{van2022spontaneous,van2024vortex}. Crystalline assemblies have been obtained in other active-matter systems both via simulations [see, e.g., Refs.~\cite{ferrante2013elasticity,menzel2014active,goto2015purely,shi2023extreme,yang2024shaping}] and experiments
 [see, e.g., Refs.~\cite{palacci2013living,tan2022odd}]; however, these crystals are not made up of spinning vortex triplets.

 In the ARCHNS model, we have identified the crucial ingredients that lead to the self-assembly of spinning vortex triplets into a triangular crystal. These ingredients -- active-crystal rotation and orientation-dependent interactions -- are already present in many experimental systems. Examples include chiral liquid crystals that show the Lehmann effect~\cite{oswald2008measurement}, biologically driven nanorods~\cite{riedel2005self}, and light-activated micromotors~\cite{palacci2013living}. Larger-scale realisations are also possible with macroscopic chiral particles propelled by airflow ~\cite{tsai2005chiral} or vibrations~\cite{deseigne2010collective}, with interactions provided by magnetic or electrostatic forces. These systems offer promising platforms for the exploration of the types of collective behaviours, which we have uncovered for $CW$ and $CCW$ active spinners;
furthermore, they suggest possibilities for controlling self-assembly in active materials. For example, programmable torque transfer~\cite{van2016spatiotemporal}, directional transport, or adaptive lattice structures may be achieved by tuning the rotation and interaction properties of the spinners. Our work contributes to the growing understanding of how activity, chirality, and interactions combine to produce novel collective dynamics in soft matter.

In a two-dimensional (2D) system in equilibrium, there can be no spontaneous symmetry breaking of a continuous symmetry~\footnote{In our system this symmetry is translational invariance.} at any finite temperature by virtue of the Hohenberg-Mermin-Wagner (HMW) theorem~\cite{hohenberg1967existence,mermin1966absence,halperin2019hohenberg}. In particular, there can be no crystalline ordering in two dimensions at any finite temperature~\cite{mermin1968crystalline,nelson1979dislocation,young1979melting,strandburg1988two} because of thermal fluctuations. This celebrated theorem does not apply to active-matter systems~\cite{menzel2014active,james2021emergence,shi2023extreme,D5SM00208G,dey2025enhanced}, so the existence of a 2D active crystal is not forbidden. However, when we do find such an active 2D crystal, it behooves us to ask how fluctuations might (or might not) eliminate strict crystalline ordering as they do in equilibrium systems. Our emergent active crystal of spinning vortex triplets provides us with a test-bed for the exploration of fundamental issues such as the formation and melting of
2D active crystals and their excitations~\footnote{For example, phonons in an equilibrium crystal.}. We will investigate these issues theoretically and by large-scale DNSs in future studies.

\section*{Model and Methods}\label{sec:Model}

 We employ the ARCHNS PDEs~\eqref{eq:phi}-\eqref{eq:incom} in which the field $\phi$ can be thought of as $\langle [\rho_{CCW}-\rho_{CW}]/[\rho_{CCW}+\rho_{CW}] \rangle$, where $\langle \cdot \rangle$ denotes a mesoscale average~\cite{cates2018theories} and $\rho_{CCW}$ and $\rho_{CW}$ are, respectively, number densities of $CCW$ and $CW$ spinners.

In two dimensions (2D), it is advantageous to use the vorticity-stream-function form~\cite{maji2025emergent}
\begin{eqnarray}
    \partial_t \omega + (\bm u \cdot \nabla) \omega &=& \nu \nabla^2 \omega -\frac{3}{2}\sigma\epsilon\nabla\times\mathcal (\nabla^2\phi\nabla\phi)  \nonumber \\
    &-&\tau\nabla^2\phi -\beta\omega \,,\label{eq:omega}
\end{eqnarray}
because the vorticity is normal to the $xy$ plane, i.e., $\bm \omega = \omega \hat{e}_z$, as is the torque $\bm{\tau} = \tau \hat{e}_z$; here, $\bm u=\nabla \times(\psi\hat e_z)$, where $\psi$ is the stream function and  $\omega = - \nabla^2 \psi$.
We can write the ARCHNS equations~\eqref{eq:phi}-\eqref{eq:incom} in the following nondimensional form:
    \begin{eqnarray}
         \partial_t \phi + (\bm u \cdot \nabla) \phi &=& \frac{3}{2Pe}\nabla^2 \left( \frac{1}{2}(\phi^3 -\phi) -Cn^2 \nabla^2 \phi \right)\,; \label{eq:phi-non} \nonumber \\
          \partial_t \omega + (\bm u \cdot \nabla) \omega &=& \frac{1}{Re} \nabla^2 \omega -\frac{3}{2}\frac{Cn}{We}\nabla\times\mathcal (\nabla^2\phi\nabla\phi) \nonumber \\ 
    &-& \alpha\nabla^2\phi  -\beta'\omega\,.\label{eq:omega-non} 
    \end{eqnarray}

\par{\textit{Initial Conditions:}} We use the following initial conditions [time $t=0$]: $\phi(x,y,t=0) = \phi_0 + \delta\phi(x,y,t=0)$, where $\phi_0$ is a space and time-independent constant and $\delta\phi(x,y,t=0)$ are independent random numbers distributed uniformly in the interval $[-0.1, 0.1]$; and $\omega(x,y,t=0) = 0$.

\par{\textit{Control Parameters, Spectra, and Spectral Balance:}} The spatiotemporal evolution of this ARCHNS system depends on the initial conditions (see above) and the following non-dimensional control parameters [see Table~\ref{tab:param}]: the Reynolds number $Re \equiv L u_{rms}/{\nu}$, the Cahn number $Cn \equiv \epsilon/L $, the Weber number $We \equiv L u_{rms}^2/\sigma$, the P\'eclet number $ Pe \equiv Lu_{rms} \epsilon/M\sigma$, and the non-dimensionalised activity $\alpha=\tau/u_{rms}^2$ and friction $\beta'=\beta L /u_{rms}$, where $u_{rms}$, $L$, and $L_{\phi}$ are, respectively, the root-mean-squared velocity, and the integral and coarsening-arrest lengths scales; the latter two are related to the energy spectrum $E(k)$ and phase-field spectrum $S(k)$ as follows: 
\begin{eqnarray}
   L&= & \frac{\sum_k k^{-1}E(k)}{\sum_k E(k)} ; \quad L_{\phi} = \frac{\sum_k S(k)}{\sum_k k S(k)}\,; \label{eq:lengths} 
\end{eqnarray}
$k$ denotes the wave number and 
\begin{eqnarray}
   E(k,t)&\equiv&\frac{1}{2}\sum_{k'= k-1/2}^{k'= k+ 1/2}|\tilde{\bm{u}}(\textbf{k}',t)|^2; \; E(k) \equiv \langle E(k,t) \rangle_t\,;\label{eq:Ek} \nonumber \\
   S(k,t)&\equiv&\frac{1}{2}\sum_{k'= k-1/2}^{k'= k+ 1/2}|\tilde{\phi}(\textbf{k}',t)|^2; \; S(k) \equiv \langle S(k,t) \rangle_t\,; \label{eq:Sk}
   \end{eqnarray}
    $k$ and $k'$ are the moduli of the wave vectors $\mathbf{k}$ and $\mathbf{k}'$; spatial Fourier transforms are denoted by tildes; and $\langle\cdot\rangle_t$ indicates the time average.
We use the integral-scale frequency $\Omega_L \equiv u_{rms}/L$ to non-dimensionalise time.



The spectral energy balance is~\cite{padhan2024novel,boffetta2012two,verma2019energy}
\begin{eqnarray}
\partial_t{{E(k,t)}} &=& - T^a(k,t) -S^{\phi}(k,t) - 2\nu k^2 E(k,t) \nonumber \\  &+& T^{b}(k,t)  - 2 \beta E(k,t)\,, \label{eq:spectralbalance}
\end{eqnarray}
 where the $T^a(k,t)$, $T^{b}(k,t)$, $S^{\phi}(k,t)$, $2\nu k^2 E(k,t)$, and $2\beta E(k,t)$ are averaged over circular shells, with radii $k$; the superscripts $a$, $b$, 
 and $\phi$ denote contributions from the advective, torque, and stress-tensor terms;
 the remaining two terms arise from viscous dissipation and friction, respectively.
 Their time averages are:
\begin{eqnarray}
    T^a(k)&=&\sum_{k'= k-1/2}^{k'= k+ 1/2} \langle\widetilde{\textbf{u}(-\textbf{k}'}).\textbf{P}(\textbf{k}').\widetilde{(\textbf{u}.    \grad\textbf{u})}(\textbf{k}')\rangle_t \,;\nonumber  \\
    T^b(k)&=&\sum_{k'= k -1/2}^{k'= k +1/2}  \langle\widetilde{\textbf{u}(-\textbf{k}'}).(\widetilde{\nabla \cp \vb{\tau}\phi})(\textbf{k}')\rangle_t\,;\nonumber  \\
    S^{\phi}(k)&=&\sum_{k'= k-1/2}^{k'= k+ 1/2} \langle\widetilde{\textbf{u}(-\textbf{k}'}).\textbf{P}(\textbf{k}').\widetilde{(\nabla^2\phi\nabla\phi)}(\textbf{k}')\rangle_t \,;\nonumber\\
    \Pi^a(k)&=&\sum_{k'=0}^{k'=k} T^a(k')\,; \;\; \Pi^b(k)=\sum_{k'=0}^{k'=k} T^b(k')\,; \label{eq:Tketc}
\end{eqnarray}
 the transverse projector $\textbf{P}(\textbf{k})$ has elements $P_{ij}(k)=\delta_{ij} -\frac{k_i.k_j}{k^2}$; and $\Pi^a(k)$ and $\Pi^b(k)$ are, respectively, the fluxes because of the advection and active stress.
The triplet-vorticity autocorrelation and cross-correlation functions (for two nearest-neighbour triplets) are, respectively, 
\begin{eqnarray}
    A(\bm{x},t) &=& \frac{1}{\mathcal{N}} \sum_{i=1}^{\mathcal{N} - t } [ (\omega(\bm{x},t_i)-\langle\omega(\bm{x},t)\rangle_t) \nonumber \\ &&(\omega(\bm{x},t_i+t) - \langle\omega(\bm{x},t)\rangle_t) ]\,\quad {\rm{and}} \nonumber \\
   C(\bm{x},t) &=& \frac{1}{\mathcal{N}} \sum_{i=1}^{\mathcal{N} - t } [(\omega(\bm{x},t_i)-\langle\omega(\bm{x},t)\rangle_t) \nonumber\\ &&
   (\omega(\bm{x}',t_i+t) - \langle\omega(\bm{x}',t)\rangle_t)]\,,
    \label{eq:timecorr}
\end{eqnarray}
where $\bm{x}'$ if the right-nearest neighbour of $\bm{x}$; for Fig.~\ref{fig:energy} (f) we choose $\bm{x}$ to be a point to the right of the triplet centre (at a distance of $\simeq 0.2731 a$), for a representative vortex triplet; 
for Fig.~\ref{fig:energy} (i) we choose $\bm{x}$  to be a point close of the triplet centre, for a representative vortex triplet.

\par{\textit{Numerical Methods:}}

We solve the ARCHNS PDEs~\eqref{eq:phi}-\eqref{eq:incom} using a pseudospectral direct numerical simulation~\cite{canuto2007spectral,padhan2024novel,padhan2025cahn} for a  square domain of size $2\pi \times 2\pi$, periodic boundary conditions in both $x$ and $y$ directions, and $N^2$ collocation points. Our DNS evaluates derivatives in Fourier space and products in physical space; it also employs the $N/2$ rule for dealiasing~\cite{canuto2007spectral,padhan2024novel,padhan2025cahn}. We use the semi-implicit ETDRK-2 method~\cite{cox2002exponential} for time marching. Our CUDA C code 
for this DNS has been developed for the NVIDIA A100 GPU processor.
We use $N=1024$, $M=0.0001$, $\epsilon=0.01839$, $\nu=0.01$, and $\beta=0.3$; the non-dimensional parameters for our DNSs are given in Table~\ref{tab:param}.

\begin{table}[h]
    \centering
\renewcommand{\arraystretch}{1.2} 
\begin{tabular}{|c|c|c|c|c|c|c|c|c|} 
 \hline
   $\tau$ & $\sigma$ & $\phi_0$ & $\alpha$ & $\beta'$  & $Cn$ & $We$ & $Pe$ & $Re$    \\ 
 \hline
  3 & 0.5 & 0.5 & 0.130 & 0.0013 & 0.266 & 2.13 & 70.78 & 38.45 \\
  \hline
  4 & 1 & 0.5 & 0.102   & 0.0009 & 0.264 &  3.44 & 90.19 &48.99   \\ 
  \hline
  4 & 2.5 & 0.5 & 0.025 &  0.0004 & 0.186 & 15.53 & 228.2 & 123.94 \\
  \hline 
  5 & 1 & 0.5 & 0.040 & 0.0004 & 0.254 & 7.19 & 132.9 & 72.19 \\
  \hline
  5 & 2 & 0.5 & 0.025 & 0.0003 &  0.214& 8.49 & 200.0 & 108.66\\
  \hline
  3 & 1 & 0.1 & 0.057 & 0.0009 & 0.166 & 7.86 & 172.0 & 93.43 \\
  \hline
  3 & 1 & 0.3 & 0.038& 0.0006 & 0.190 & 10.23 & 183.3 & 99.57 \\
  \hline
  4 & 1 & 0.3 & 0.023 & 0.0003 & 0.207 & 15.50 & 216.3 & 117.49 \\
  \hline
  4 & 1 & 0.4 & 0.023 & 0.0003 & 0.233 & 13.58 & 190.8 & 103.66 \\
  \hline
\end{tabular}
\caption{\textbf{Table of the values of the non-dimensional parameters, for different values of $\tau,\,\sigma$, and $\phi_0$ [columns 1, 2, and 3]:} the non-dimensionalised activity $\alpha=\tau/u_{rms}^2$ and friction $\beta'=\beta L /u_{rms}$, where $u_{rms}$ is the root-mean-square velocity and $L$ is the integral length scale [Eq.~\eqref{eq:lengths}]; the Cahn number $Cn \equiv \epsilon/L $; the Weber number $We \equiv L u_{rms}^2/\sigma$; the P\'eclet number $ Pe \equiv Lu_{rms} \epsilon/M\sigma$; and the Reynolds number $Re \equiv L u_{rms}/\nu$.}
 \label{tab:param}
\end{table}
\begin{acknowledgments}
		We thank A. Jayakumar, K.V. Kiran, E. Knobloch, and  S. Ramaswamy for useful discussions. B. Maji and R. Pandit thank the Anusandhan National Research Foundation (ANRF), the Science and Engineering Research Board (SERB), and the National Supercomputing Mission (NSM), India, for support,  and the Supercomputer Education and Research Centre (IISc), for computational resources. The authors would like to thank the Isaac Newton Institute for Mathematical Sciences, Cambridge, for support and hospitality during the programme \textbf{Anti-diffusive dynamics: from sub-cellular to astrophysical scales}, where some of the work on this paper was undertaken; this work was supported by EPSRC grant EP/R014604/1. NB Padhan was supported by the German Research Foundation (DFG) through the project ``Analysing structure-property relations in equilibrium and non-equilibrium hyperuniform systems'' (project number VO 899/32-1).
	\end{acknowledgments}
  \section*{Appendix}
  \subsection*{Videos}
  The following videos show the spatiotemporal evolution of the pseudocolor plots in Fig ~\ref{fig:schematic}, ~\ref{fig:vort_diagram}, ~\ref{fig:field_diagram}:
  \begin{itemize}
\item Video V0: This shows the spatiotemporal evolution of pseudocolor renderings [cf. Figs. 1 (e)] of the fields  $\omega$  for $\tau=4$ and $\sigma=1.0$, $\phi_0=0.5$ (\textcolor{blue}{\href{https://youtu.be/-zlqWCohmVw}{https://youtu.be/-zlqWCohmVw}});
\item Video V1: This shows the spatiotemporal evolution of pseudocolor renderings [cf. Figs. 2 (b)] of the fields  $\omega$  for $\tau=3$ and $\sigma=0.5$, $\phi_0=0.5$ (\textcolor{blue}{\href{https://youtu.be/yW3YIuxmWLA}{https://youtu.be/yW3YIuxmWLA}});
\item Video V2: This shows the spatiotemporal evolution of pseudocolor renderings [cf. Figs. 2 (d)] of the fields  $\omega$  for $\tau=5$ and $\sigma=1$, $\phi_0=0.5$\\ (\textcolor{blue}{\url{https://youtu.be/Ol2qByaxxsM}});
\item Video V3: This shows the spatiotemporal evolution of pseudocolor renderings [cf. Figs. 2 (f)] of the fields  $\omega$  for $\tau=4$ and $\sigma=2.5$, $\phi_0=0.5$\\ (\textcolor{blue}{\url{https://youtu.be/ahXQzyaCsxs}});
\item Video V4: This shows the spatiotemporal evolution of pseudocolor renderings [cf. Figs. 2 (h)] of the fields  $\omega$  for $\tau=5$ and $\sigma=2$, $\phi_0=0.5$ \\(\textcolor{blue}{\url{https://youtu.be/Z7rtr2tkOVw}});
\item Video V5: This shows the spatiotemporal evolution of pseudocolor renderings [cf. Figs. 3 (b)] of the fields  $\omega$  for $\tau=3$ and $\sigma=1$, $\phi_0=0.1$ \\(\textcolor{blue}{\url{https://youtu.be/L2PJwb3uxGE}});
\item Video V6: This shows the spatiotemporal evolution of pseudocolor renderings [cf. Figs. 3 (d)] of the fields  $\omega$  for $\tau=4$ and $\sigma=1$, $\phi_0=0.3$ \\(\textcolor{blue}{\url{https://youtu.be/Y0DLfzc8log}});
\item Video V7: This shows the spatiotemporal evolution of pseudocolor renderings [cf. Figs. 3 (f)] of the fields  $\omega$  for $\tau=3$ and $\sigma=1$, $\phi_0=0.3$\\ (\textcolor{blue}{\url{https://youtu.be/gCYhfWixrYI}});
\item Video V8: This shows the spatiotemporal evolution of pseudocolor renderings [cf. Figs. 3 (h)] of the fields  $\omega$  for $\tau=4$ and $\sigma=1$, $\phi_0=0.4$ \\(\textcolor{blue}{\url{https://youtu.be/ux7ttqzN-PM}}).
\end{itemize}
\newpage
\bibliography{main}

\end{document}